\documentclass[a4paper]{article}

\usepackage{19lomcon}        
\usepackage{cite}             
\usepackage{epsfig}           
\usepackage{epstopdf}
\usepackage{graphicx,epsf,amsmath,amsfonts,amssymb,amsbsy}
\newcommand{\vev}[1]{\langle#1\rangle}
\newcommand{\ds}{\displaystyle}
\newcommand{\mat}{\left ( \begin{array}}
\newcommand{\emat}{\end{array} \right )}
\newcommand{\vect}{\left ( \begin{array}{c}}
\newcommand{\evect}{\end{array} \right )}

\begin{document}


\title{
Quark/hadronic matter and dualities of QCD thermodynamics 
}

\author{T. G. Khunjua $^{1,2}$, K. G. Klimenko $^{3}$ 
,
        R. N. Zhokhov $^{3,4}$ \email{zhokhovr@gmail.com}
}

\affiliation{
$^{1}$ \quad Faculty of Physics, M.V. Lomonosov Moscow State University, Moscow\\
$^{2}$ \quad University of Georgia\\
$^{3}$ \quad Logunov Institute for High Energy Physics, NRC "Kurchatov Institute", Protvino, Moscow Region \\
$^{4}$ \quad  Pushkov Institute of Terrestrial Magnetism, Ionosphere and Radiowave Propagation (IZMIRAN), Troitsk, Moscow;}


\date{}
\maketitle


\begin{abstract}
  In this letter the dualities of thermodynamics of dense quark/hadronic matter are discussed. It is shown that chiral symmetry breaking (CSB) phenomenon is dual to 
  charged pion condensation (PC) phenomenon.
  . 
\end{abstract}

\section{Introduction}


It is believed that strong interation phenomena described by quantum chromodynamics (QCD). QCD is notoriously hard to deal with especially at finite baryon density where even lattice QCD approach faces insurmountable difficulties. 
One of the ways out is to use effective models such as NJL model. Many different physical quantities are studied in terms of this model. Resemblant four-fermion models are also discussed in low dimensions as a toy model for QCD \cite{Schon:2000qy, Winstel:2019zfn, kkzz} or even 
in connection to condensed matter systems \cite{Yoshii:2019yln, caldas}. In this letter the so-called duality properties of QCD phase diagram are discussed. They are the dualities of 
QCD thermodynamics 
(of the phase structure of QCD) with different densities (such as baryon density, chirality, isospin asymmetry). 

\section{The model and matter with non-zero isospin and chiral densities}

Let us discuss here a phase structure of the two flavored dense quark matter with different imbalances.
In order to describe non-zero baryon (quark) density 
$n_B\sim n_{u}+n_{d}\neq0$ one need to add to the Lagrangian the following term $\frac{\mu_B}{3}\bar q\gamma^0q$.
The isospin imbalance $n_{u}-n_{d}\neq0$ can be introduced as 
$\frac{\mu_I}2\bar q \tau_3\gamma^0q$.
The more exotic condition, chiral  $n_{5}=n_{R}-n_{L}\neq0$ and chiral isospin $
n_{5}^{u}-n_{5}^{d}\neq0$ imbalances can be accounted for by the following terms $\mu_5\bar q\gamma^{0}\gamma^{5}q$ and $\frac{\mu_{I5}}2\bar q \tau_3\gamma^0\gamma^5q$.

It is considered in (3+1)-dimensional NJL model which Lagrangian has the form
\begin{eqnarray}
L=\bar q\Big [\gamma^\nu\mathrm{i}\partial_\nu
+\frac{\mu_B}{3}\gamma^0+\frac{\mu_I}2 \tau_3\gamma^0+\frac{\mu_{I5}}2\tau_{3} \gamma^0\gamma^5+\mu_{5} \gamma^0\gamma^5\Big ]q+ \frac
{G}{N_c}\Big [(\bar qq)^2+(\bar q\mathrm{i}\gamma^5\vec\tau q)^2 \Big
],  \label{1}
\end{eqnarray}
where $q$ field is the flavor doublet, $q=(q_u,q_d)^T$, where $q_u$ and $q_d$ are four-component Dirac spinors. 
$\tau_k$ ($k=1,2,3$) are Pauli matrices. 
From now on the notations
$\mu\equiv\mu_B/3$, $\nu\equiv\mu_I/2$, $\nu_{5}\equiv\mu_{I5}/2$ will be used.

To find the thermodynamic potential (TDP) of the system, we use a semi-bosonized version of the Lagrangian,
\begin{eqnarray}
\widetilde L\ds = \bar q\Big [\gamma^\rho\mathrm{i}\partial_\rho
+\mu\gamma^0+\nu\tau_3\gamma^0+\nu_{5}\tau_{3}\gamma^0\gamma^5+\mu_{5}\gamma^0\gamma^5-\sigma -\mathrm{i}\gamma^5\pi_a\tau_a\Big ]q
 -\frac{N_c\Big (\sigma^2
 +\pi_a^2
 \Big )}{4G},
\label{2}
\end{eqnarray}
 which contains composite bosonic fields $\sigma (x)$ and $\pi_a (x)$ 
 with equations of motion
\begin{equation}
\sigma(x)=-2\frac G{N_c}(\bar qq);~~~\pi_a (x)=-2\frac G{N_c}(\bar q
\mathrm{i}\gamma^5\tau_a q).
\label{200}
\end{equation}
Note that the composite bosonic field $\sigma$ can be identified with sigma meson (or $f_{0}(500)$) \cite{Pelaez:2015qba},  whereas bosonic field $\pi_3 (x)$ can be identified with the physical $\pi^0(x)$-meson and fields $\pi_{1,2} (x)$ with the physical charged $\pi^\pm (x)$-mesons by means of the following expression $\pi^\pm (x)=(\pi_1 (x)\mp i\pi_2 (x))/\sqrt{2}$.

Here we suppose that the ground state expectation values $\vev{\sigma(x)}$ and $\vev{\pi_a(x)}$ do not depend on $x$ and
without loss of generality one can 
the following ansatz
\begin{equation}
\vev{\sigma(x)}=M,~~~\vev{\pi_1(x)}=\Delta,~~~\vev{\pi_2(x)}=0,~~~ \vev{\pi_3(x)}=0. \label{06}
\end{equation}
In the leading order of the large-$N_c$ expansion TDP can be shown to have the following form 

\begin{equation}
\Omega (M,\Delta)=
f(M^2+\Delta^2, a_+, b_+, c_+)\times f(M^2+\Delta^2,a_-, b_-, c_-), \label{5}
\end{equation}
where

 $a_\pm=M^2+\Delta^2+(|\vec p|\pm\mu_{5})^2+\nu^2+\nu_{5}^2;~~b_\pm=\pm 8(|\vec p|\pm\mu_{5})\nu\nu_{5};\nonumber\\
c_\pm=a_\pm^2-4 \nu ^2
\left(M^2+(|\vec p|\pm\mu_{5})^2\right)-4 \nu_{5}^2 \left(\Delta ^2+(|\vec p|\pm\mu_{5})^2\right)-4\nu^{2} \nu_{5}^2.$

\section{
Duality relations.}

It can be seen 
 from Eq (\ref{5}) that the TDP is invariant with respect to the transformation
\begin{equation}
{\cal D}:~~~~M\longleftrightarrow \Delta,~~\nu\longleftrightarrow\nu_5.
 \label{16}
\end{equation}
It is a so-called main duality 
and it tells us that we can simultaneously exchange chiral condensate and charged pion condensate and isospin and chiral imbalances and the results do not change. This means that CSB phenomenon in the system with isospin (chiral) imbalance is the same as (equivalent to) charged PC phenomenon in the system with chiral (isospin) imbalance. It is an interesting property of the phase structure.


Besides the main duality 
there are two other so-called constrained dualities, which include additional constraint.
For example, it is possible to show that with the constraint $\Delta=0$ (if there is no charged PC
 ) the TDP (\ref{5}) is invariant with respect to the following transformation
\begin{equation}
{\cal D}_M:~~~~\Delta=0,~~~\mu_5\longleftrightarrow\nu_{5},
 \label{19}
\end{equation}
meaning that CSB phenomenon does not feel the differece between two types of chiral imbalance (chiral and chiral isospin imbalances).

Also one can demonstrate that 
with the constraint $M=0$ (chiral symmetry is 
restored
) the TDP is invariant with respect to the transformation
\begin{equation}
{\cal D}_\Delta:~~~~M=0,~~~\mu_5\longleftrightarrow\nu.
 \label{21}
\end{equation}
This means that charged PC phenomenon is influenced in the same way by completely different imbalances of the system, namely chiral and isospin ones. 

Moreover, let us note that the dualities ${\cal D}_M$ and ${\cal D}_\Delta$ are dual to each other with respect to the main duality ${\cal D}$.
The sketch clarifying the dualities and their interrelations
is depicted in Fig. 1.
\begin{figure}
\centering
\includegraphics[width=0.7\textwidth]{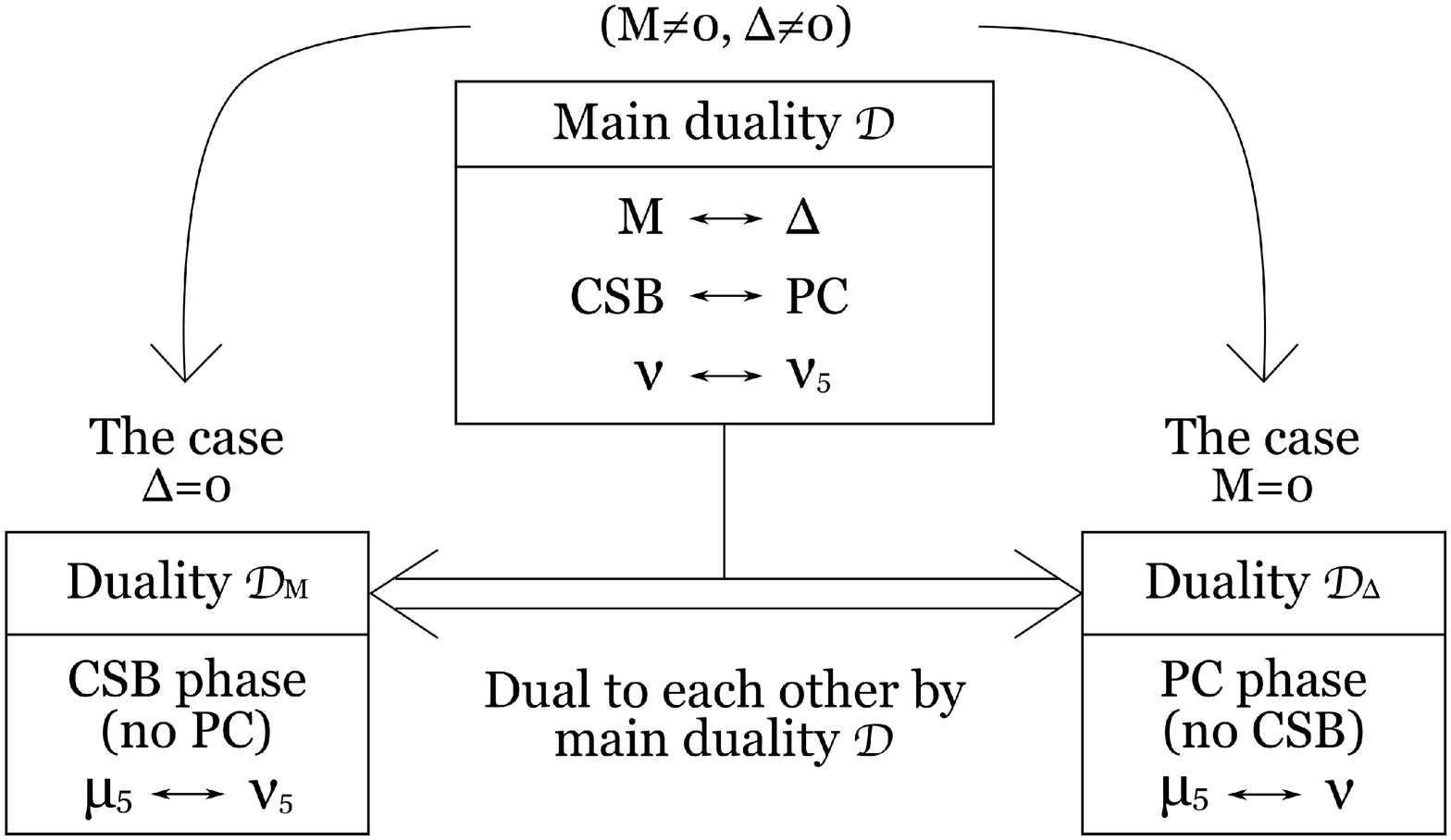}
\caption{The main duality  $\cal D$ is valid in the most general case, 
the dualities  ${\cal D}_{M}$ and ${\cal D}_\Delta$ valid only 
outside of PC and CSB phases.}
\end{figure}

\section*{Acknowledgments}

R.N.Z. is grateful for support of Russian Science Foundation under the grant No 19-72-00077.
The work is also supported by the Foundation for the Advancement of Theoretical Physics and Mathematics
BASIS grant.


\end{document}